\documentclass[twocolumn,10pt]{asme2e}

\usepackage{mathrsfs}
\usepackage{bbm}
\usepackage{comment}
\usepackage{amsmath}
\usepackage{cite}
\usepackage{amsfonts}
\usepackage{amssymb,latexsym}
\usepackage[psamsfonts]{eucal}
\usepackage{graphicx}
\usepackage{epsfig}
\usepackage{psfrag}
\usepackage{subfigure}
\usepackage{pstcol}
\usepackage{epsf}
\usepackage{epstopdf}
\usepackage{amssymb}
\usepackage{amsmath}
\usepackage{ulem}
\usepackage{color,soul}

\confshortname{IDETC/CIE 2018}
\conffullname{the ASME 2018 International Design Engineering Technical Conferences \&\\
              Computers and Information in Engineering Conference}

\confdate{26-29}
\confmonth{August}

\confyear{2018}
\confcity{Quebec City}
\confcountry{Canada}

\papernum{DETC2018-85449}

\title{FOLLOWER FORCES IN PRE-STRESSED FIXED-FIXED RODS TO MIMIC OSCILLATORY BEATING OF ACTIVE FILAMENTS}

\author{Soheil Fatehiboroujeni
    \affiliation{
	Department of Mechanical Engineering\\
	University of California\\
	Merced, California 95343\\
    Email: sfatehiboroujeni@ucmerced.edu
    }	
}

\author{Arvind Gopinath
    \affiliation{
	Department of Bioengineering\\
	University of California\\
	Merced, California 95343\\
    Email: agopinath@ucmerced.edu
    }	
}

\author{Sachin Goyal
    \affiliation{
	Department of Mechanical Engineering\\
	Health Science Research Institute\\
	University of California\\
	Merced, California 95343\\
    Email: sachin.goyal@ucmerced.edu
    }	
}

\begin{document}

\setlength{\abovedisplayskip}{0.2cm}
\setlength{\belowdisplayskip}{0.2cm}

%{\red Red color: comments from Sachin}
%
%{\blue Blue color: Soheil's follow up}
%
%{\red
%\begin{enumerate}
%  \item Check if references are really supposed to come before appendices or after. 
%  {\blue
%- references are supposed to come before the appendices}
%  \item Does the inverse method uses exact knowledge of the boundary condition?
%\end{enumerate}
%}
%{\blue
%
%-ASME convention: "An example
%is shown in Eqn. (1). The number of a referenced equation in
%the text should be preceded by Eqn. unless the reference starts a
%sentence in which case Eqn. should be expanded to Equation."
%}
\maketitle

\begin{abstract}
  {Flagella and cilia are examples of actively oscillating, whiplike biological filaments that are crucial to processes as diverse as locomotion, mucus clearance, embryogenesis and cell motility. Elastic driven rod-like filaments subjected to compressive follower forces provide a way to mimic oscillatory beating in synthetic settings. In the continuum limit, this spatiotemporal response is an emergent phenomenon resulting from the interplay between the structural elastic instability of the slender rods subjected to the non-conservative follower forces, geometric constraints that control the onset of this instability, and viscous dissipation due to fluid drag by ambient media. In this paper, we use an elastic rod model to characterize beating frequencies, the critical follower forces and the non-linear rod shapes, for pre-stressed, clamped rods subject to two types of fluid drag forces, namely, linear Stokes drag and non-linear Morrison drag. We find that the critical follower force depends strongly on the initial slack and weakly on the nature of the drag force. The emergent frequencies however, depend strongly on both the extent of pre-stress as well as the nature of the fluid drag.}  
\end{abstract}

\begin{comment}

\begin{nomenclature}
\entry{A}{You may include nomenclature here.}
\entry{$\alpha$}{There are two arguments for each entry of the nomenclature environment, the symbol and the definition.}
\end{nomenclature}

The spacing between abstract and the text heading is two line spaces.  The primary text heading is  boldface in all capitals, flushed left with the left margin.  The spacing between the  text and the heading is also two line spaces.

\end{comment}

\section{INTRODUCTION}
\label{sec:intro}

\begin{figure*}[t]
\centering
\includegraphics[width=1.86\columnwidth]{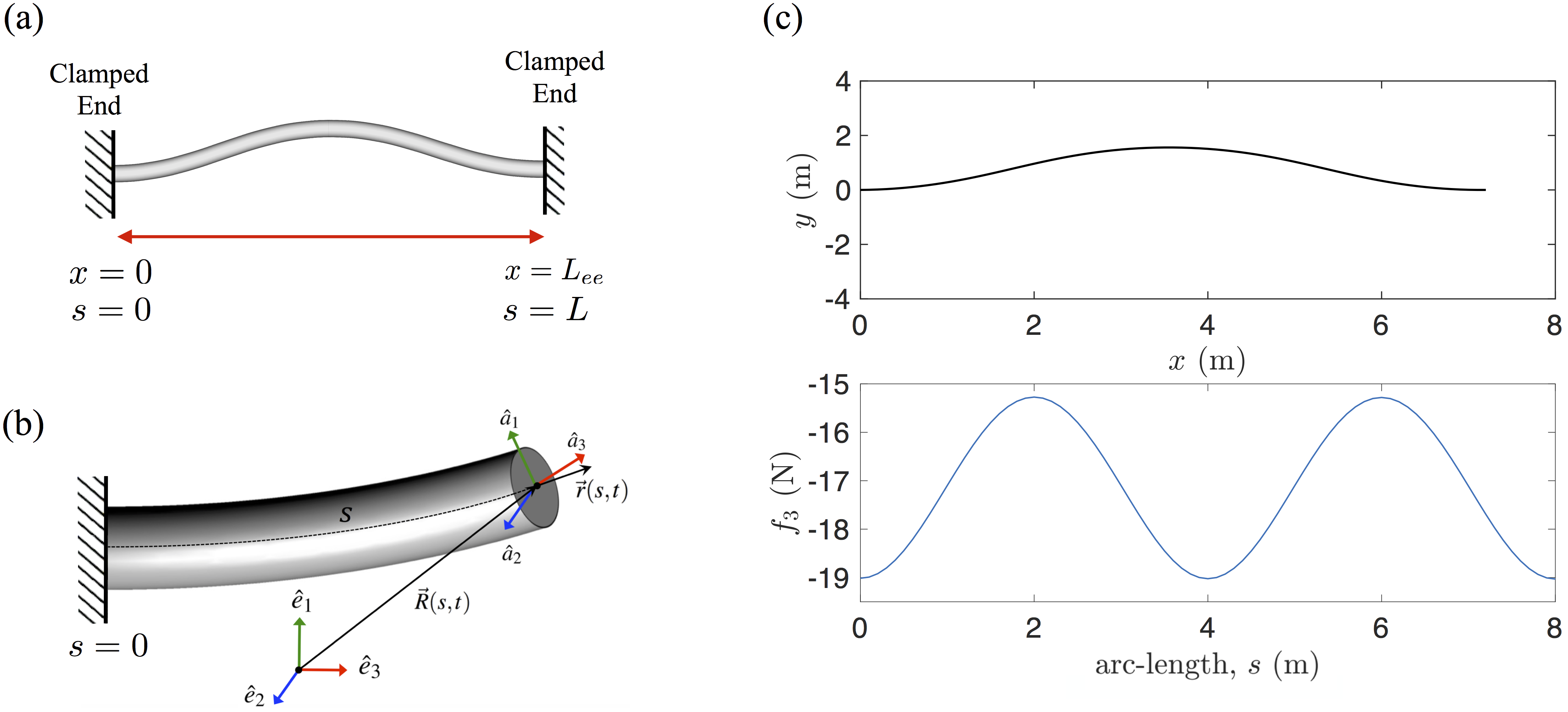}
\caption{(a) Schematic representation of a rod of unstressed length $L$ with fixed-fixed boundary condition (clamped at both ends). The end to end distance when buckled is $L_{\mathrm{ee}} < L$. (b) The motion of material points comprising the cross-section of the rod at arc-length position, $s$ and at time $t$ is determined by tracking the transformations of the body-fixed frame $\hat{a}_{i}(s,t)$ with respect to the inertial frame of reference $\hat{e}_{i}$. (c) The shape (top) and pre-stress (bottom) in the buckled state with $L_{\mathrm{ee}}/L = 0.9$. Pre-stress is determined by the component of the Internal force in the direction of cross-sectional normal vector $\hat{a}_{3}(s,t)$ i.e., $f_3$.}
\label{fig:schem_1}
\end{figure*}

Stability analysis of slender structures subjected to follower loads are important instantiations of nonconservative problems in the theory of elastic stability. A number of thorough surveys of the developments and achievements on the structural stability of nonconservative systems can be found in the literature \cite{LANGTHJEM2000809, Elishakoff2005, Bolotin99}.
Conservative loads such as gravitational or electrostatic forces can be written as gradient of a time independent potential function \cite{Leipholz01}. 

The elastic buckling of a straight, slender column subject to compressive forces is a classical problem involving conservative forces that has received much attention. Consider an Euler column of length $L$, modulus of elasticity $E$ and moment of inertia $I$, clamped at one end and subject to a concentrated axial load $P$ at the other. Let the column deform transversely with a (dimensional) lateral displacement $h$ as a result of this forcing. Since the force is conservative, we can write a potential function
$V=\frac{1}{2}EI\int_0^\ell h''\,^2 ds - \frac{1}{2}P\int_0^L h'\,^2 ds$ using which variational techniques deliver the shape of the buckled column $h''' + (P / EI)  h' = 0$. 
The critical load $P_{\mathrm{cr}}$ for instability and subsequent buckling is obtained by seeking solutions to this equation subject to boundary conditions $h(0)=h'(0)=h''(L)=0$. The critical load  $P_{\mathrm{cr}}=\left(EI\pi^2/4L^2\right)$ can be then determined as the minimum loading that makes the potential energy $V$ cease being positive definite.  However, such static linear stability analyses are applicable only to systems where the loading forces are not time dependent.  

Nonconservative loads, however, do not fit this criteria; their magnitude and direction depend on the configuration of a structure (e.g., deflection and slope), its velocity, and time. An ubiquitous example of a nonconservative force 
is fluid drag and viscous damping forces  that depend on the velocity of a structure. Follower forces are a second type of nonconservative force; these, acting either as a point force at one end or a distributed load along the structure, always act tangential to the deflection curve of a structure. Reut \cite{Reut01}, Pfluger \cite{Pfluger01}, and Beck \cite{Beck1952} were among the first researchers to analyze the buckling of cantilevers subjected to follower forces. For instance, Beck \cite{Beck1952} reports that the critical buckling load for a cantilever subjected to a compressive point load that always remains tangential to the free-end of the cantilever is approximately 10 times larger than for a force with constant direction.  
Stability analysis of slender structures subjected to follower loads is critical in several applications including pipes conveying fluid \cite{Paidoussis2016, PaIdoussis1993}, structures which are self propelled by their own thrust \cite{Wood1969}, and rockets \cite{Paidoussis1973}. It is shown that motion equations of disc-brake systems \cite{Kang2000} is identical to that of Leipholz's column which is a cantilever subjected to a compressive and uniformly distributed follower load. 

More recently, follower forces have been studied in the context of connected actively driven beads. Synthetic filaments comprised of connected beads when actuated provide a mechanism to mimic the oscillatory beating of flagella and cilia \cite{Dreyfus2005, Babataheri2011, Qin2015}.  
While the length scales are much smaller in these applications (ranging from around 1-500 $\mu$m), dynamical principles underlying their structural stability remain similar; indeed connections between mechanics at multiple length scales have been illustrated in other biological settings \cite{Vaziri2008, Gopinath2011, Vaziri2007}. 
For example, tunable colloidal chains that are assemblies of Janus particles with controllable polarities can be tuned to generate oscillatory beating \cite{Sasaki2014, PATTESON201686, PMID:24352670} by subjecting them to tangential compressive (follower) forces. 
Continuum models are used in numerous studies and have shown to be effective tools for analyzing the post-buckling behavior of rods subjected to follower force. Dissipation of energy due to viscous drag plays an important role in determining the steady-state beating frequency as well as the critical value of the follower load \cite{DeCanio20170491, Herrmann1964, PMID:24352670, Bayly2016}. 

Previous studies of filaments subject to follower forces--both for viscous drag dominated as well as inertial filaments--have focused on mobility and dynamics of free-free, fixed-free, and pinned-free filaments with the base state being a straight non-stressed filament or rod. The role of pre-stress in emergent oscillations driven by active distributed follower forces is yet to be elucidated. Here, we focus on this complementary scenario of a fixed-fixed rod.  Specifically, a rod clamped at both ends is first pre-stressed by decreasing the end-to-end distance, thereby generating a buckled shape and then subjected to follower force. In the fixed-free scenario, the lack of constraint at the free-end allow for either lateral oscillations or steady rotations to develop in favorable conditions \cite{PMID:24352670}. In the fixed-fixed scenario, the slack generated upon initial compression  (which may alternately be interpreted as a pre-stress) offers the necessary degree of freedom to allow for oscillations. Our simulations are three-dimensional, but, by introducing strictly planar perturbations and loads, the oscillations remain planar. 

Note that in general, stability of systems subject to non-conservative forces involve linear operators that is not self-adjoint. Critical buckling loads in such non self-adjoint systems cannot be determined using Euler's static method but rather by estimating eigenvalues of the associated dynamic problem. Alternatively, critical points and post-buckling solutions may be obtained by solving the fully non-linear, time-dependent problem as done in this work.

\section{MODEL}
\label{sec:rod}

\begin{table}[h]
\begin{center}
{\small
    \begin{tabular}{ | l | l | l | l |}
    \hline
    {\bf Quantity} & {\bf Variable} & {\bf Value} & {\bf Units} \\ \hline\hline
    Diameter & $d$ & $ 0.0096 $ & m \\ \hline
    Length & $L$ & 8 & m \\ \hline
    Mass per unit length& $m$ & 0.2019 & kg/m \\ \hline
    Young's modulus & $E$ & 68.95 & GPa \\ \hline
    Shear modulus & $G$ & 27.58 & GPa \\ \hline
    Second moment of area & $I_1=I_2$ & 4.24 $\times$10$^{-10}$ & m$^4$ \\ \hline
    Polar moment of area & $I_3$ & 8.48 $\times$10$^{-10}$ & m$^4$ \\ \hline
    Normal drag coefficient & $C_n$ & 0.1 & m.s or m$^2$ \\ \hline
    Tangential drag coefficient & $C_t$ & 0.01 & m.s or m$^2$ \\ \hline
    Surrounding fluid density & $\rho_{\textrm{f}} $ & 1000 & kg/m$^3$  \\ \hline
    \end{tabular}
    \caption{Numerical values for the geometric and elastic properties of the rod and drag coefficients used in the computations. }
    \label{tab1}}
\end{center}
\end{table}

We consider a rod that is in its stress-free state when maintaining a straight shape. By moving one of the clamped ends of the rod toward the other end and forcing the rod to bend due to bucking as shown in Fig 1(a), we generate pre-stress in the rod. Thus, pre-stress rate is controlled by the end-to-end length of the rod, $L_{ee} < L$.  

\subsection{Governing equations}

The continuum rod model that we use \cite{goyal:05b} follows the classical approach of the Kirchhoff
\cite{Kirchhoff} which assumes each cross-section of the rod to be rigid. The rigid-body motion of individual cross-sections is examined by discretizing an elastic rod into infinitesimal elements along its arc-length. The position and orientation of each cross-section is determined in space $s$ (i.e., the arc-length variable) and time $t$ by tracking the transformation
of a body-fixed frame $\hat{a}_i(s, t)$ with respect to an inertial frame of reference $\hat{e}_i(s, t)$ as shown in Fig 2(b), where subscript $i = 1,2,3$.

Vector $\vec{R}(s, t)$ defines the position of the cross-sections relative to the inertial frame of reference. The spatial derivative of $\vec{R}(s, t)$ is denoted by vector $\vec{r}(s, t)$. Deviation of $\vec{r}(s, t)$ from the unit 
normal of the cross-section determines shear while the change in its magnitude quantifies stretch (extension or compression) along the arc-length $s$. Both shear and stretch deformations are negligible for filaments with large slenderness (length/thickness) ratio under compression. So, we assume $\vec{r}(s, t)=\hat{a}_3 (s, t)=\hat{t}(s, t)$, where $\hat{t}(s, t)$ is the unit tangent vector along the arc-length. Vector $\vec{\kappa}(s,t)$ captures two-axes bending and torsion of the rod and vectors $\vec{v}(s,t)$ and $\vec{\omega}(s,t)$ represent the translational velocity and the angular velocity of cross-sections, respectively. The stress distribution over the cross-section of the rod results in a net internal force and 
a net internal moment shown respectively with $\vec{f}(s,t)$ and $\vec{q}(s,t)$.

The equations of equilibrium (\ref{linear_momentum}) and (\ref{angular_momentum}) are derived by applying Newton's second law to an infinitesimal element of the rod. The compatibility equations (\ref{position_continuity}) and (\ref{orient_continuity}) follow from the space-time continuity of the cross-section position $\vec{R}(s,t)$, and the space-time continuity of the transformation from $\hat{a}_i(s, t)$ to $\hat{e}_i(s, t)$. In Eqns. (\ref{linear_momentum})-(\ref{orient_continuity}) all derivatives are relative to the body-fixed reference frame, $m$ is the mass of the rod per 
unit length and $\underline{\mathbf{I_m}}$ is a 3-by-3 tensor of the moments of inertia per unit length. External force per unit length $\vec{F}$ as well as the external moment per unit length $\vec{Q}$ capture interactions of the rod with itself \cite{goyal:08c, lillian2011electrostatics} or the environment such as drag force.
\begin{eqnarray}
m(\frac{\partial \vec{v}}{\partial t} + \vec{\omega} \times \vec{v}) - (\frac{\partial \vec{f}}{\partial s} + \vec{\kappa} \times \vec{f}) - \vec{F} &=& \vec{0} \label{linear_momentum}\\
\underline{\mathbf{I_m}}\frac{\partial \vec{\omega}}{\partial t} + \vec{\omega} \times {{ \underline{\mathbf{I_m}}}}\vec{\omega} -(\frac{\partial \vec{q}}{\partial s} + \vec{\kappa} \times \vec{q}) + \vec{f} \times \vec{r} - \vec{Q}&=& \vec{0} \label{angular_momentum}\\
\frac{\partial \vec{r}}{\partial t} + \vec{\omega} \times \vec{r} - (\frac{\partial \vec{v}}{\partial s} + \vec{\kappa} \times \vec{v})  &=&\vec{0} \label{position_continuity}\\
\frac{\partial \vec{\kappa}}{\partial t} - (\frac{\partial \vec{\omega}}{\partial s} + \vec{\kappa} \times \vec{\omega})  &=&\vec{0}  \label{orient_continuity}
\end{eqnarray}

The distributed follower forces and moments in this model are captured by $\vec{F}$ and $\vec{Q}$. In the scenario of fixed-fixed rod, we consider the effect of distributed follower forces in tangential direction (along $\hat{a}_3(s, t)$) in this paper. Hereforward for simplicity of notation, we refer to this tangential follower force density by scalar $F$. In the scenario of fixed-free rods, there may also be point follower loads at one of the boundaries. 

The differential equations of equilibrium and compatibility (\ref{linear_momentum})-(\ref{orient_continuity})
have to be solved together with a constitutive law relating the deformations to the restoring forces. 
The constitutive law, for an elastic rod may, in general, take the form of an implicit algebraic constraint \cite{hinkle}, but for an isotropic and linearly elastic rod that is not curved in the stress-free state \cite{goyal:07a} it takes an explicit form \cite{goyal:08b}:
\begin{eqnarray}
 \vec{q}(s,t) &=& \underline{\mathbf{B}}(s) \vec{\kappa}(s,t). \label{constitutive_law}
\end{eqnarray}
The matrix $\underline{\mathbf{B}}$ in equation (\ref{constitutive_law}) encodes the bending and torsional stiffness moduli 
of the rod. By choosing the body-fixed frames of reference to coincide with principal axes of 
rod cross-sections, $\underline{\mathbf{B}}$ can be written as
\begin{eqnarray}
   \underline{\mathbf{B}}=
  \left[ {\begin{array}{ccc}
    EI_1 & 0 & 0 \\
    0 & EI_2 & 0 \\
    0 & 0 & GI_3\\
  \end{array} } \right],
  \label{stiffness}
\end{eqnarray}
where $E$ is the Young's modulus, $G$ is the shear modulus, and $I_1$, $I_2$, and $I_3$ 
represent the second moment of area of the rod cross-section about its principal axes. Our choice, as implicit in Fig 1(b),  implies 
that subscripts $i = 1,2$ in $\hat{a}_i(s, t)$ represent the rod's axes of bending and $i = 3$ represents torsional axis.

 \subsection{Solution scheme}
 
The \textit{Generalized-$\alpha$} method \cite{chung:93a} is adopted to compute
the numerical solution of this system, subjected to necessary
and sufficient initial and boundary conditions. A detailed description of 
this numerical scheme applied to this formulation is given in the extant literature \cite{goyal:06a}.  The important feature of this method is that it is an unconditionally stable second order accurate method for numerically stiff problems, which allows for controllable numerical dissipation. In the context of rod mechanics, it brings an improvement over box method \cite{gobat:01a} by controlling the Crank-Nicolson noise, in which numerical solution oscillates about the true solution at every (temporal or spatial step) and corrupts the subsequent computation. While in the analysis presented here, the constitutive equations are linear and local, as embodied in equations (5) and (6), the method can be adapted to analyze problems where the constitutive relationships are non-linear and non-local.

\begin{figure}
\centering
\includegraphics[width=0.86\columnwidth]{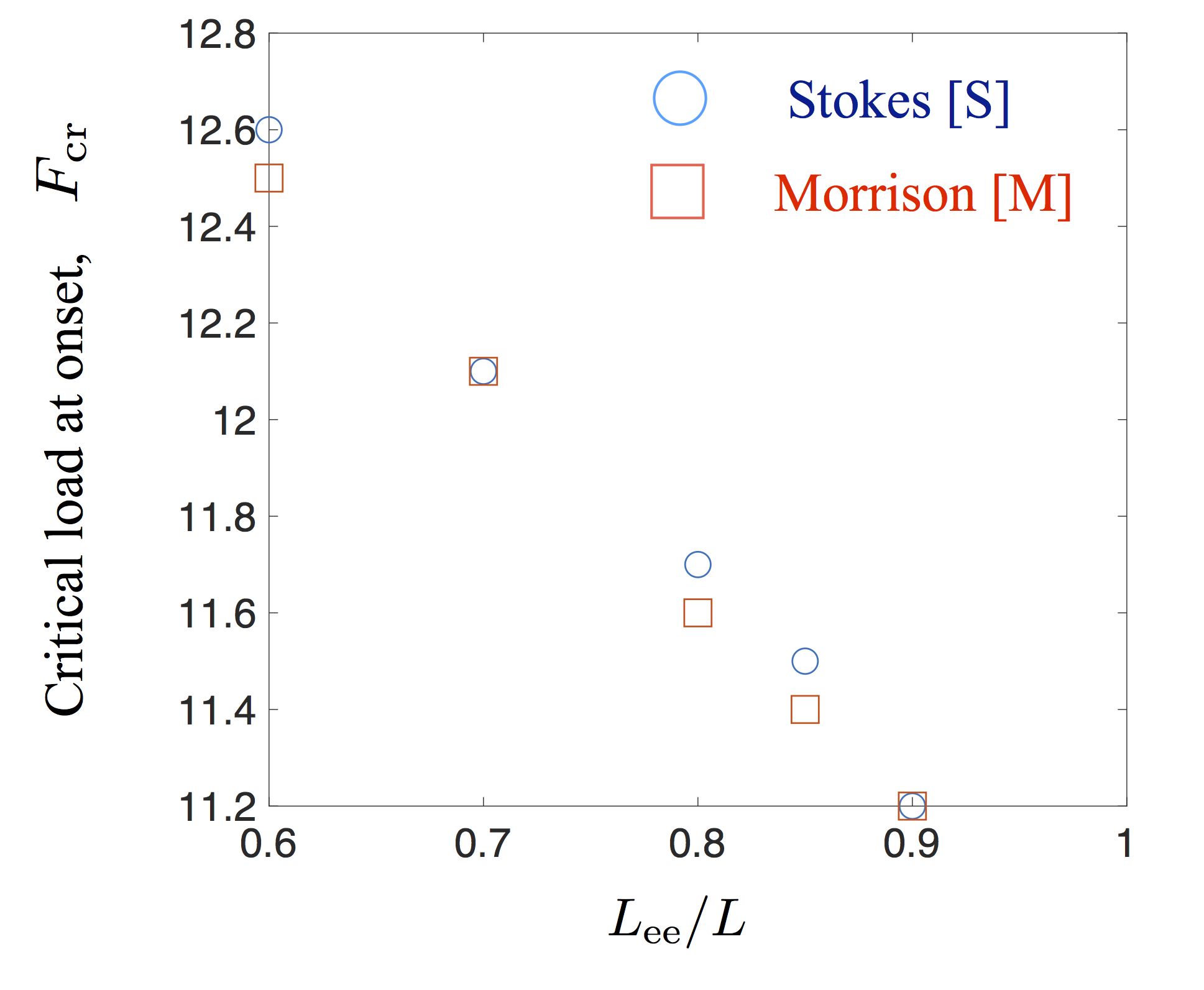}
\caption{Critical load for onset of oscillations $F_{\mathrm{cr}}$ versus scaled end-to-end distance $L_{\mathrm{ee}}/L$ for both Stokes [S] drag and Morrison [M] drag. We note that the critical loads are roughly the same over the range of pre-stress values investigated.}
\label{fig:FL}
\end{figure}

\begin{figure}
\centering
\includegraphics[width=\columnwidth]{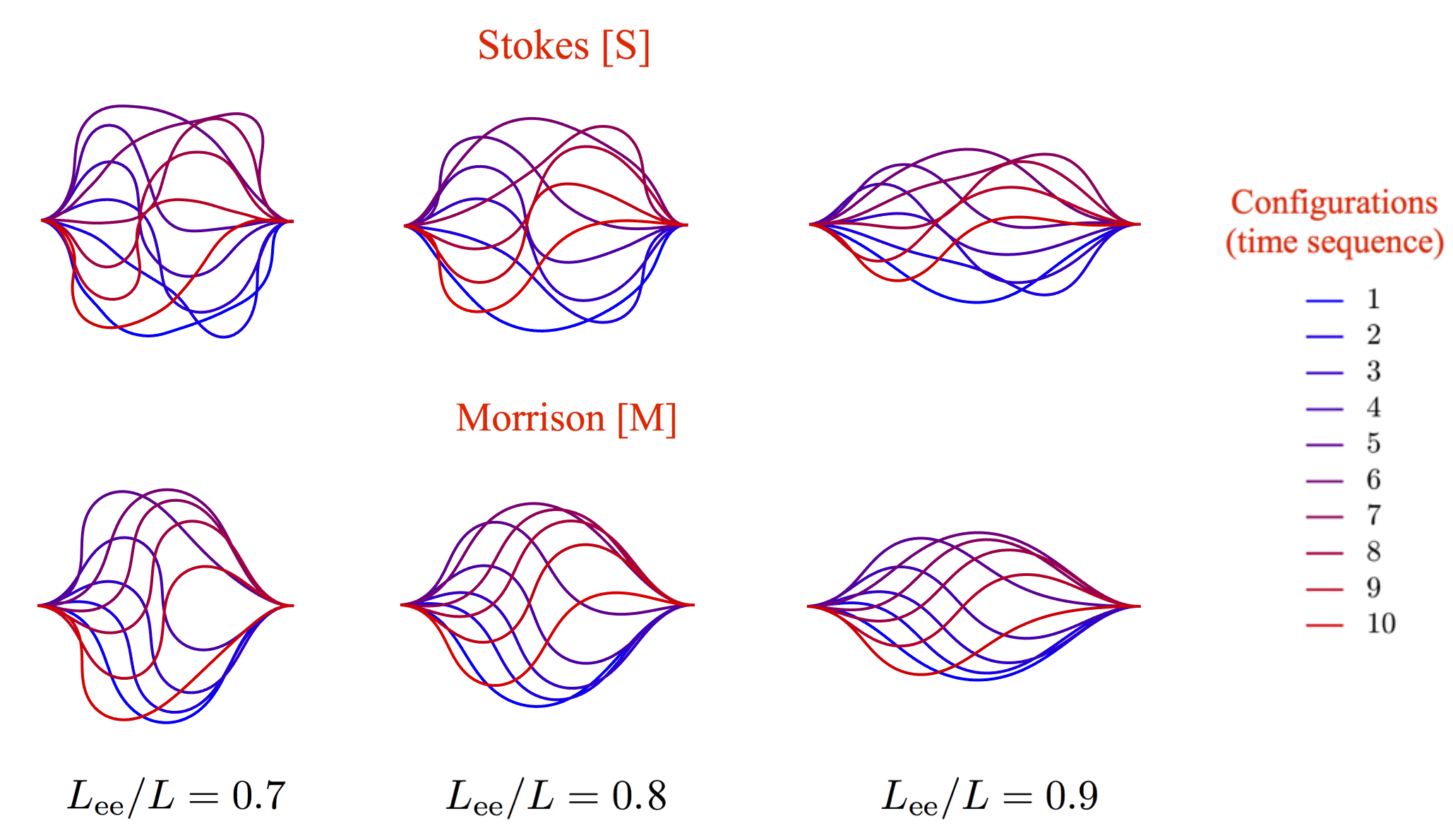}
\caption{Configurations of the rod in sequence (1-10) over one period of oscillation when $|F|=15$ {N}/{m} and for different values of the dimensionless slack $L-L_{\mathrm{ee}}/L$. The configurations (rod shapes) for Stokes drag are shown on the top, with shapes for  Morrison drag shown in the bottom row.}
\label{fig:shapes}
\end{figure}

\section{RESULTS}
\label{sec:results}

We next present results for the critical value of the follower force density $F_{\mathrm{cr}}$ versus end-to-end distance $L_{\mathrm{ee}}/L$ and explore how the  beating frequency, $\omega(|F|, L_{\mathrm{ee}}/L)$ both at the critical point and for values of the follower force $|F| > F_{\mathrm{cr}}$ depends on the pre-stress. A cylindrical rod with slenderness ratio of 800 is simulated with properties given in Table \ref{tab1}. We compare the findings for two types of drag forces, namely Stokes (S) drag and Morrison (M) drag forms given in Eqs. (\ref{eq:dragS}) and (\ref{eq:dragM}), respectively as explained in \cite{goyal:05b}. 
\begin{equation}
\vec{F}_{\textrm{S}}=-\frac{1}{2}\rho_{\textrm{f}} d \Big( C_n \: \hat{t}\times(\vec{v}\times \hat{t}) + \pi \:C_t(\vec{v}\cdot \hat{t})\:\hat{t} \Big)
\label{eq:dragS}
\end{equation}
\begin{equation}
\vec{F}_{\textrm{M}}=-\frac{1}{2}\rho_{\textrm{f}} d \Big( C_n|\vec{v}\times \hat{t}|\: \hat{t}\times(\vec{v}\times \hat{t}) + \pi \:C_t(\vec{v}\cdot \hat{t})|\vec{v}\times \hat{t}|\:\hat{t} \Big)
\label{eq:dragM}
\end{equation}
In both equations $\rho_{\textrm{f}}$ and $d$ represent the environment fluid density and diameter of the rod, respectively. Drag coefficients (per unit length) $C_n$ and $C_t$ are given in Table \ref{tab1}. We note that the Stokes [S] form for the drag is linear in the velocity while the Morrison form [M] is quadratic, and hence non-linear in the velocity.
Thus for the same change in configuration and frequency, the Morrison form will result in a larger viscous dissipation per unit length than the Stokes form. Conversely, if we require that the same amount of energy be dissipated, the Stokes limit will be characterized by either higher frequency or by larger amplitude deformations or both. 

\begin{figure*}
\centering
\includegraphics[width=2\columnwidth]{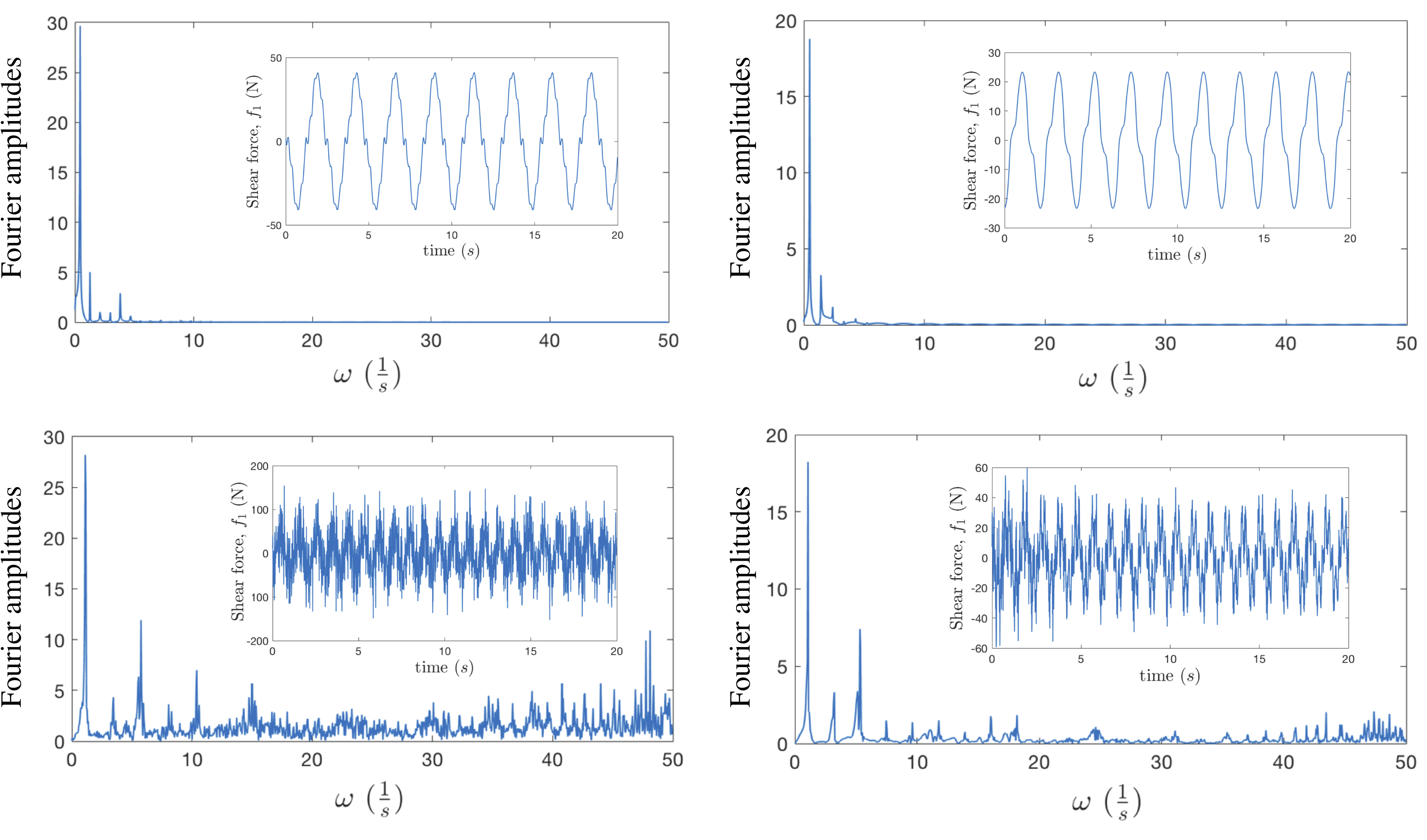}
\caption{Fourier transform (in the time domain) of the shear force at the mid-span length of the rod shows that higher harmonics (insets show the raw data) are damped in the case of the non-linear Morrison drag (top row) more effectively than for the linear Stokes drag (bottom row). The right column of the picture corresponds to $L_{\mathrm{ee}}/L=0.9$ while the left column corresponds to $L_{\mathrm{ee}}/L=0.7$. The ratio of drag coefficients for both cases is 10. Intuitively, we expect this ratio to affect the extent of dampening. }
\label{fig:FFTs}
\end{figure*}

\begin{figure*}[t]
\centering
\includegraphics[width=2\columnwidth]{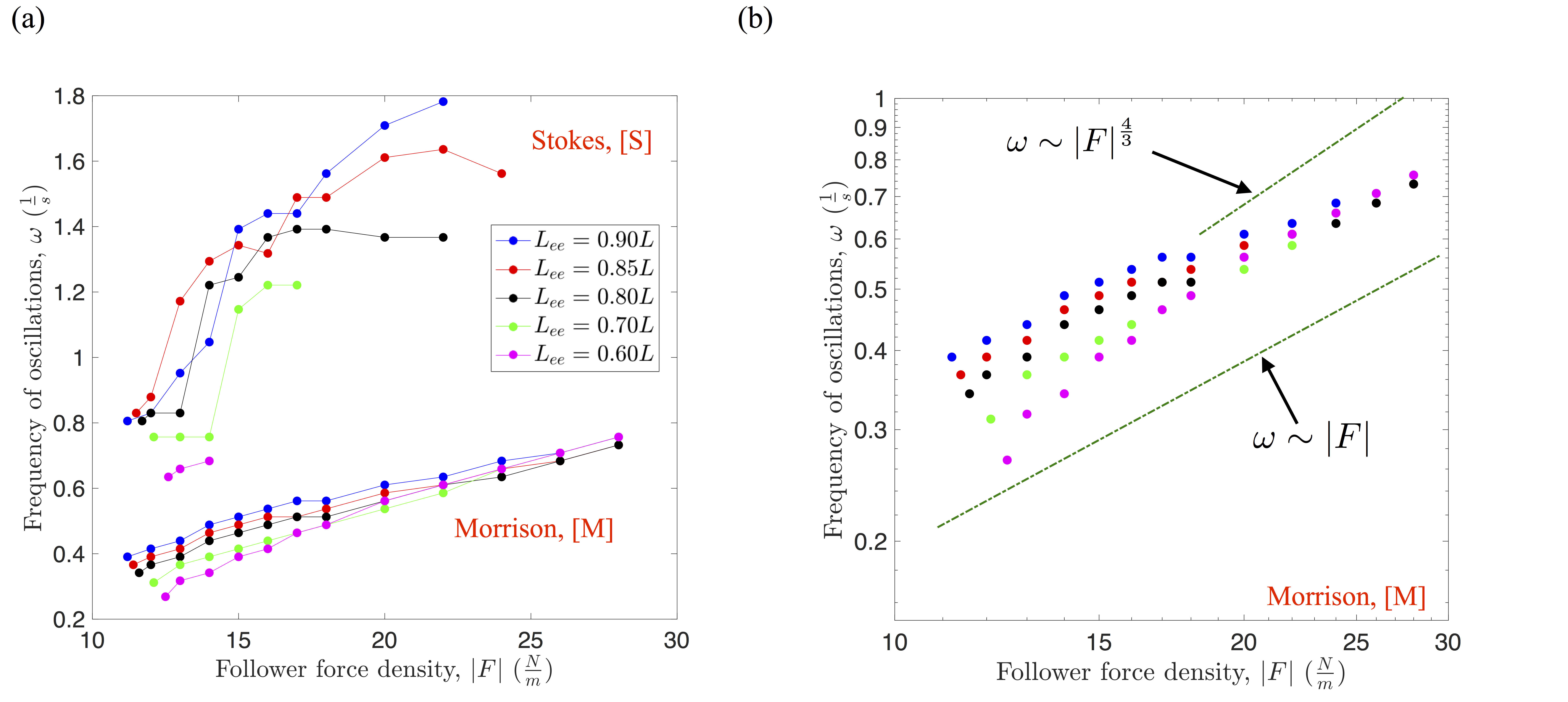}
\caption{(a) Frequency of beating oscillations for rods having various end-to-end distances $L_{\mathrm{ee}}$. We show results for both types of fluid drag - the linear Stokes [S] drag as well as the quadratic  Morrison [M] drag.  The frequency is plotted as a function of the distributed follower load. We note that the results for the Stokes drag features possible transitions that may be related to activation of higher order mode shapes as seen from figure 3. (b) Results for the case where the viscous drag of of the Morrison [M] form are re-plotted in logarithmic scales to illustrate two salient features - (i) as the follower force increases to values much larger than the critical values, the effect of the pre-stress diminishes, and (ii) the frequencies in the limit $|F| \gg F_{\mathrm{cr}}$ vary as $|F|^{\alpha}$ where the exponent $1 < \alpha < 4/3$.} 
\label{fig:freq}
\end{figure*}

\subsection{Benchmark: Critical Force for Beck's Column}
In order to benchmark the model presented in this paper, we calculate the critical buckling force for the Beck's column and compare it with the values reported. Beck's column is a cantilever subjected to a compressive point load that is always tangential to the free-end of the column. Beck's analysis, published in German \cite{Beck1952} and reviewed in English \cite{Elishakoff2005} yields the following expression for the critical buckling force, $P_{\mathrm{cr}}$ of a cantilever with bending stiffness, $EI$ and length, $L$ in absence of damping dissipation. $
P_{\mathrm{cr}} \approx 20.51 \: ({EI}/{L^2})$.

Using the formulation presented in Section \ref{sec:rod}, we investigate the value of the critical buckling force for Beck's column and compare it to the value reported in the literature. To approach the conditions of a quasi-static simulation and reduce the dynamic effects we apply a compressive follower force, which gradually increases in time, to the free-end of the cantilever. The critical force found by our computational model in absence of viscous drag is approximately $ P_{\mathrm{cr}} \approx 20.10 \:({EI}/{L^2})$, which is within two percent error margin of the classical estimated value. 

\subsection{Oscillatory Beating of Fixed-Fixed Rods}
In this section we present the results for the post-buckling analysis of pre-stressed rods with fixed-fixed boundary conditions for various values of the the slack (and thus, various values of the pre-stress as well as base curvature). Identifying and characterizing critical points as well as the force-frequency relationship can be potentially useful in designing accurately controllable oscillations. 

We study cylindrical rods that are in stress-free state when straight. When both ends of a rod are fixed and clamped, moving one of the clamped ends of the rod toward the other end forces the rod to bend or buckle (c.f., Figure 1(a)). This process generates pre-stress in a rod and thus, pre-stress rates can be controlled by the end-to-end length of the rod, $L_{ee}$ - an example of this is shown in Fig. 1(c). Starting from a base state completely determined by the ratio $L_{\mathrm{ee}}/L$, we then apply uniformly distributed follower load, $F\hat{a}_3$ along the rod. When the magnitude of the follower load, $|F|>F_{\mathrm{cr}}$, buckled shapes become unstable and beating oscillations emerge. Figure \ref{fig:FL} shows the magnitudes of the critical follower load against end-to-end distance for the same rod subjected to two types of drag forces. Surprisingly we observe that critical follower load increases as the amount of pre-stress in the rod increases even though decreasing $L_{ee}/{L}$ implies more slack. The magnitude of critical follower load found to be nearly the same for both Stokes and Morrison drags (discrepancies being $\pm 0.1$\%). 

Despite the low sensitivity of $F_{\mathrm{cr}}$ to the nature of drag law, beating configurations as well as the steady frequency of oscillations are found to be significantly different for Stokes and Morrison drags. This can be explained by the fact that Morrison drag dissipates energy at a higher rate compared to the Stokes drag for the same frequency and mode shapes. Figure \ref{fig:shapes} illustrates how the shape of the rod evolves during one complete oscillation for both Stokes and Morrison drags. We visually observe that configurations of the rod subjected to Stokes drag consist of various shape modes. This suggests that higher order harmonics are stronger in this case. Whereas, for cases subjected to Morrison drag, higher order shapes are not recognizable visually. By looking at the Fourier transformation of any quantity we can also confirm the significance of higher order harmonics in Stokes regime compared to Morrison as is shown in Figure \ref{fig:FFTs}.

Finally, with the computational model proposed here we systematically investigate the effect of pre-stress and the follower force on the frequency of beating oscillations and emergent shapes. It is useful, at this point, to recall previous results on unstressed fixed-free cantilever type rods subject to follower forces and Stokes drag  \cite{PMID:24352670} with equal values of axial and normal drag. As in our simulations, fixed-free unstressed rods were found to undergo an instability to oscillatory motion beyond a critical value of the follower force; in this case, the lack of pre-stress implies that the instability occurs at a single critical point $F_{\mathrm{cr}} \approx 75.5 (EI/L^{3})$ (where we recall that the follower force is a force density with units of N/m) .
Interestingly relaxing the boundary condition to a pinned-free rod was found to alter the dynamical picture completely. In the latter case the post-bucked state was not an oscillating filament but instead a rotating spiral. In either case, a single dimensionless parameter $\beta \equiv FL^{3}/EI$ encodes the role of activity in the post-buckling shapes and frequencies post criticality. For  $\beta > 75.5$, the frequency $\Omega$ was found to {\em monotonically increase} with $\beta$ (and hence with $|F|$) and eventually follow a power law relationship $\Omega  \sim |F|^{4 \over 3}$ for constant filament properties.
For extremely high values of $|F|$, self-avoidance resulted in a change in this scaling to a different power law with exponent of 2.
 
Moving now to our results for pre-stressed filaments, we plot the frequency of beating oscillations for rods under various end-to-end distances
in figures \ref{fig:freq}(a,b).  We observe that frequency of oscillations under Stokes drag undergoes a sudden increase once the magnitude of the distributed follower load reaches a \textit{second critical limit}. Such a behavior is absent under Morrison drag. Also for a region where magnitude of the distributed follower load is below 18 N/m we observe that larger pre-stress (or smaller end-to-end distance) results in smaller beating frequency. This pattern is also evident under the Stokes drag but only in the region in which the magnitude of the distributed follower load is above 18 N/m. Examining the force dependence of the beating filaments subject to Morrison type drag forces more closely in Figure 5(b), we find the frequencies in the limit $|F| \gg F_{\mathrm{cr}}$ vary as $|F|^{\alpha}$ where the exponent is closer to unity than to 4/3. More interestingly, we note also the weakening dependence of the frequency on the pre-stress (slack) in the limit of large follower force densities - that is when $|F|/F_{\mathrm{cr}} \gg 1$. This behavior is particularly evident in Figure 5(b).

\section{DISCUSSION}
\label{sec:discussion}
In this paper we discussed the application of a computational rod model to analyze the buckling stability as well as the post-buckling oscillations of slender structures subjected to compressive follower loads. Simulations were first benchmarked with previous findings on magnitude of the critical buckling force for Beck's column. We focused on slender rods that maintain a straight shape corresponding to their stress-free state (i.e., having no intrinsic curvature and twist) with both ends clamped. By moving one end of the rod toward the other end, the structure undergoes buckling and the end-to-end distance represents a measure of the amount of pre-stress in the rod. We found that beyond a critical value of distributed and compressive follower loads the buckled shapes become unstable and oscillatory beating emerges. The magnitude of the critical follower load increases as the magnitude of the pre-stress in the structure increases. We also observed that frequency of the oscillations as well as the configuration of the rod are significantly influenced by the type of drag law used in modeling. Morrison drag induces higher dissipation rate than Stokes drag, therefore, under identical circumstances many more harmonics are discernible in the oscillations of a rod subjected to Stokes drag. Moreover, for the rods subjected to Stokes drag we observed that frequency of oscillations as a function of follower load undergoes a sudden increase once the magnitude of the distributed follower load reaches a \textit{second critical limit}.  Our results provide a starting point to systematically investigate the interplay between geometry, elasticity, dissipation and activity towards designing bio-inspired multi-functional, synthetic structures to  move and manipulate fluid at various length scales.

%{\color{red} Soheil - some of the reference in the bibliography seem incomplete/not standard notation - can you check again? Thanks!}

\bibliographystyle{asmems4}

\bibliography{asme2e}

\end{document}